\documentclass[pra,superscriptaddress,twocolumn
,floatfix
]{revtex4-1}

\usepackage{hyperref}
\usepackage{braket}
\usepackage{graphicx}
\usepackage{amsmath}
\usepackage{times,dsfont,amssymb,amsmath,amsthm,amsfonts,amsbsy,mathrsfs,bm,bbm,graphicx,graphics,epsfig,multirow,mathtools,color,bbold,empheq}
\usepackage{complexity}

\graphicspath{{./}}

\newclass{\SampP}{SampP}
\newclass{\SampBQP}{SampBQP}
\newclass{\IQP}{IQP}
\newclass{\BosonSampling}{BosonSampling}
\newclass{\FBPP}{FBPP}

\newcommand{\LON}{\hbox{LON}}

\newcommand{\ketbra}[1]{\ket{#1}\hspace{-0.3em}\bra{#1}}

\newcommand{\midx}{\bm{m}}
\newcommand{\nidx}{\bm{n}}
\newcommand{\Usub}{U_{1_N \times 1_N}}
\newcommand{\Usm}{U_{1_N \times \nidx}}
\newcommand{\Uprob}[1]{|\text{Per} \left(#1\right)|^2}
\newcommand{\Usubprob}{\Uprob{\Usub}}

\begin{document}

\title{Exact BosonSampling using Gaussian continuous variable measurements}

\newcommand{\affqct}{Centre for Quantum Computation and Communications
	Technology, School of Mathematics and Physics, The University of
	Queensland, St Lucia, Queensland, 4072, Australia}
\newcommand{\affipm}{School of Physics, Institute for Research in Fundamental Sciences (IPM),
	P.O.Box 19395-5531, Tehran, Iran}

\author{A.~P.~Lund}
\affiliation{\affqct}
\author{S.~Rahimi-Keshari}
\affiliation{\affipm}
\affiliation{\affqct}
\author{T.~C.~Ralph}
\affiliation{\affqct}

\begin{abstract}
	$\BosonSampling$ is a quantum mechanical task involving Fock
	basis state preparation and detection and evolution using only linear
	interactions.  A classical algorithm for producing samples from this
	quantum task cannot be efficient unless the polynomial hierarchy of
	complexity classes collapses, a situation believe to be highly
	implausible.  We present method for constructing a device which uses
	Fock state preparations, linear interactions and Gaussian
	continuous-variable measurements for which one can show exact sampling
	would be hard for a classical algorithm in the same way as Boson
	Sampling.  The detection events used from this arrangement does not
	allow a similar conclusion for the classical hardness of
	approximate sampling to be drawn.  We discuss the details of this
	result outlining some specific properties that approximate sampling
	hardness requires.
\end{abstract}

\maketitle

\section{Introduction}

$\BosonSampling$ is the task of producing statistical samples from Fock basis
measurements of a bosonic $M$-mode linear scattering network with an input
consisting of $N$ modes prepared with a single boson and the remaining
$M-N$ modes prepared in the vacuum state. This task, whilst not
universal for quantum computing, has been shown to not be efficiently
computable by any classical algorithm (or the polynomial hierarchy of
complexity classes collapses which is believe to be extremely
unlikely)~\cite{AA}.  However, a quantum implementation is efficient as one
merely needs to build the scattering device as described within a sufficiently
small error budget.  This is many orders of magnitude easier than the
construction of a fully universal quantum computer, but is still a challenge
for current technology.

Attempts have been made to identify scenarios where the proof of classical
hardness of $\BosonSampling$ can be used or adapted to other sampling
problems.  One particular scenario that is of experimental interest is in the
use of continuous variable (CV) Gaussian states or measurements.  For another
restricted computational model based on sampling from qubit circuits involving
commuting coherent rotations it has been shown that CV variants are hard to
simulate classically~\cite{Douce17}.  It is known that for linear networks
with Gaussian state inputs and Gaussian measurements it is efficient for
classical algorithms to not only produce samples but compute the entire output
distribution~\cite{Bartlett03}.   Never-the-less it has been shown that a
hybrid approach which involves linear networks with input two-mode squeezed
vacuum states and Fock basis detection has a similar classical hardness proof
to the original $\BosonSampling$ problem~\cite{Lund2014}. There is also
evidence for the classical hardness of a more general construction involving
squeezed-vacuum inputs, linear optics and Fock basis
detection~\cite{Rahimi2015, Ham2016}.  The question this paper addresses is
the reverse situation, Fock state inputs to linear networks and homodyne
detection.

An important aspect of the hardness proof for $\BosonSampling$ is that the
output probability distribution contains probabilities which are proportional
to matrix permanents from sub-matricies of the matrix describing the linear
scattering network.  A matrix permanent is a quantity which is computed like a
matrix determinant without the alternating addition and subtraction.  In fact,
when sampling with Fock state inputs and detection, all detection
probabilities are proportional to sub-matricies derived from the linear
scattering network~\cite{Scheel04}.  This special situation, given two
plausible conjectures hold, allows for the hardness proof of ``approximate''
sampling to be shown~\cite{AA}.  This is because an allowed error budget's
effect can be spread over all detection events provided the linear network
appears randomly distributed.

Here we show that using single-photon input states and a particular Gaussian
measurement one can extract sub-matrix permanents to within an exponentially
small error.  Therefore one can show that exact sampling from this
distribution is hard.  It is not necessarily the case that approximate
sampling is still hard and we discuss this in relation to our construction.

In section II we present some of the background behind the hardness arguments
for $\BosonSampling$.  Then in section III we present the CV-n detector model
using a Fock state $\ket{n}$ input with CV measurements.  In section IV we
will then describe exact sampling using the CV-1 model and the technical
details involved in showing the hardness of computing samples from the CV
output distribution.  Finally we will discuss the issues preventing the
hardness result from being used in this model to make definitive
conclusions about the hardness or not of approximate sampling.

\section{Classical hardness of $\BosonSampling$}

A problem in the class $\BosonSampling$ is one where the statistical samples
can be generated by an $M$ mode linear interaction between $N$ singly occupied
bosonic modes (and $M-N$ bosonic vacua) which is subsequently detected in the
Fock basis.  $\BosonSampling$ is either inefficient using classical
computational resources (not in $\P$), or the ``polynomial hierarchy'' of
complexity classes collapses to the third level, a situation believe to be
implausible.  It is not our goal to present in full the background and
subsequent arguments towards the truth of this statement as this has been done
elsewhere~\cite{AA}.  We will, however, outline some of the key aspects used
in this paper that are needed to understanding what is required for the proof
presented in~\cite{AA} to hold.

\subsection{Polynomial hierarchy}

The polynomial hierarchy of complexity classes is a nested structure defined
by the use of oracles.  An oracle is essentially an assumption on the
resources available to an algorithm which can greatly assist in proving
statements in computational complexity.  The hierarchy has a complex
definition and we will concentrate on a simplified version. 

The class $\NP$ is the set of decision ('yes'/'no') problems whose satisfying
input can be verified efficiently.   This defines the first level of the
polynomial hierarchy.  The second level is then the class $\NP$ with access to
an oracle from the first level.  Subsequent levels are defined by continuing
this recursion, e.g. the third level is the class $\NP$ with access to an oracle
from the second level.

This structure has a strong connection to similarly defined hierarchies within
number theory and set theory.  In those cases each level of the hierarchy is
strictly larger than lower levels.  If two levels were to coincide, then the
addition of levels stops growing the hierarchy is said to collapse.  In terms
of the computational complexity structure, a collapse of the hierarchy to the
first level means that $\P = \NP$ or that $\NP$ problems can be efficiently
solved deterministically a situation believed to be highly implausible.  A
collapse to the second level would mean $\P^{\NP} = \NP^{\NP}$ which is the
same statement relative to an $\NP$ oracle.  Being relative to the oracle
means that the statement is slightly more plausible, but it is still believed
to be not possible.  The prevailing belief is that the polynomial hierarchy
does not collapse to any level and this is the assumption on which the
hardness of $\BosonSampling$ can be proven. 

\subsection{Stockmeyer's approximate counting algorithm}

Critical to the hardness proof of $\BosonSampling$ is the use of Stockmeyer's
approximate counting algorithm~\cite{Stockmeyer1983}.  This algorithm computes
estimates of a quantity defined as
\begin{equation}
	F = \sum_{x \in Q}f(x)
\end{equation}
where $f:Q \rightarrow \{0,1\}$ is a boolean function from length $l$
bitstrings $Q = \{0,1\}^l$ onto a single bit.   In other words $F$ is the
number of inputs that result in an output of $1$, a set we will call $Q_1$.
The computed estimate of this quantity is multiplicative, which means the
estimate of $F$ that the algorithm produces is $\tilde{F}$ satisfying
\begin{equation}
	F g^{-1} \leq \tilde{F} \leq  F g
\end{equation}
where $g>1$ and for Stockmeyer's algorithm is lower bounded by $1 + 1/poly(l)$.

Stockmeyer's algorithm computes the estimate $\hat{F}$ by finding the smallest
output with no collisions for a randomly chosen hash functions on $Q_1$.  That
is, choose randomly a function $h: Q_1 \rightarrow \{0,1\}^p$ (for $p \leq l$)
and if there exists no elements $a,b \in Q_1$ such that $h(a) = h(b)$ (a hash
collision) then we have made an upper bounded estimate on the size of $Q_1$
of $2^p$.   

Here we can see the elements which make up this algorithm: the estimate of size
is multiplicative, it requires finding hash collisions, which is an $\NP$
problem, and involves random choices.  This results in Stockmeyer's algorithm
being contained within a class $\FBPP^\NP$ (if the function $f$ is efficiently
computable).  The superscript notation describes access to an $\NP$ oracle,
which in the case of Sockmeyer's algorithm is used to find hash collisions.
$\BPP$ means the algorithm proceeds by making random (probabilistic) choices
with a probability of success at least $2/3$.  The prefix 'F' is used to
describe the output from this algorithm, which is a number (function) rather
than a decision or 'yes'/'no' output.   This is therefore not an algorithm one
would expect to be efficiently computable when realistically implemented.  But
it is used here to establish if a problem lives within the ``polynomial''
hierarchy of complexity classes.  In this case, this algorithm lives within
the third level as $\BPP \subset \NP^{\NP}$~\cite{Sip,Laut}. 

This algorithm is used to make multiplicative estimations of the underlying
probabilities of a distribution from samples of the distribution.  In the
model of classical computation, unlike quantum mechanical models, there is no
inherent randomness.  Randomness is introduced by external means and can be
regarded as an input to the algorithm.   If the function $f(x)$ above
represent a sampling algorithm for a two-outcome distribution, then it can be
thought of as converting input random bits $x$ distributed uniformly into
samples of the desired probability distribution.   One can then estimate the
probability of the outcome $1$ by counting how many inputs produce the outcome
$1$ relative to the total number of possible inputs.  This would mean dividing
the estimate $\hat{F}$ by $2^l$ which will also produce a multiplicative
estimate even though it has been divided by an exponentially increasing
factor. 

A polynomial hierarchy collapse is triggered if $f$ is efficient to compute and
the probability it produces samples from is a quantity for which multiplicative
estimation is known to be outside the third level of the polynomial hierarchy.

\subsection{Multiplicative and additive errors}

Producing approximations within multiplicative errors is particularly powerful.
It is more natural to consider the case of additive errors.  An additive
estimate of $F$ would be one satisfying
\begin{equation}
	F - g \leq \tilde{F} \leq F + g.
\end{equation}
This kind of estimation will arise from more natural models of errors within a
computation.  For example, the function $f$ above most likely admits errors
which are of an additive nature and hence the estimations performed using this
noisy model will also be additive.  Only in the case of admitting exactly zero
noise to $f$, or in the sampling case a situation called ``exact sampling'',
can one generate a multiplicative estimation.

The crucial outcome of~\cite{AA} is that the polynomial hierarchy collapse can
be shown to occur in $\BosonSampling$ even with a given level of total
variation distance between the ideal and actual distributions.   The total
variation distance is an additive quantity and will generate additive errors
in estimates.  But as the probabilities in $\BosonSampling$ tend to decrease
to zero exponentially, the introduction of additive errors will overwhelm the
magnitude of the quantity being estimated.  The trick of~\cite{AA} is to
use the structure of the probabilities in $\BosonSampling$ to encode the
estimated quantity within an exponentially large set of possible outcomes in a
way that without knowledge of where the problem is encoded, looks like a
Haar-random unitary matrix (i.e. the choice is hidden from the
implementation).  This means that the additive estimate generated will have
bounds determined by the average case which turns out to be sufficient to
additively estimate matrix permanents.  Using this estimate it is then
possible to trigger a polynomial hierarchy collapse to the third level, given
that two plausible conjectures about the nature of estimating permanents of
Gaussian random matrices hold true. 

\section{The detector}

We will now describe our continuous-variable detection model and how we use it
to construct a probability distribution which can be used in the arguments
of~\cite{AA}. The model is based on a measurement device for measuring in the
displaced number state basis, which we refer to it as CV-n measurement, and
two variations of this measurement, phase-randomised CV-n (PRCV-n) and
discretised-phase-randomised CV-n (DPRCV-n).  Only a brief outline of the
model is presented here.  The details of this calculation are given in
Appendix~\ref{app:cvn}.

\begin{figure}[h]
	\centering
	\includegraphics[width=1\columnwidth]{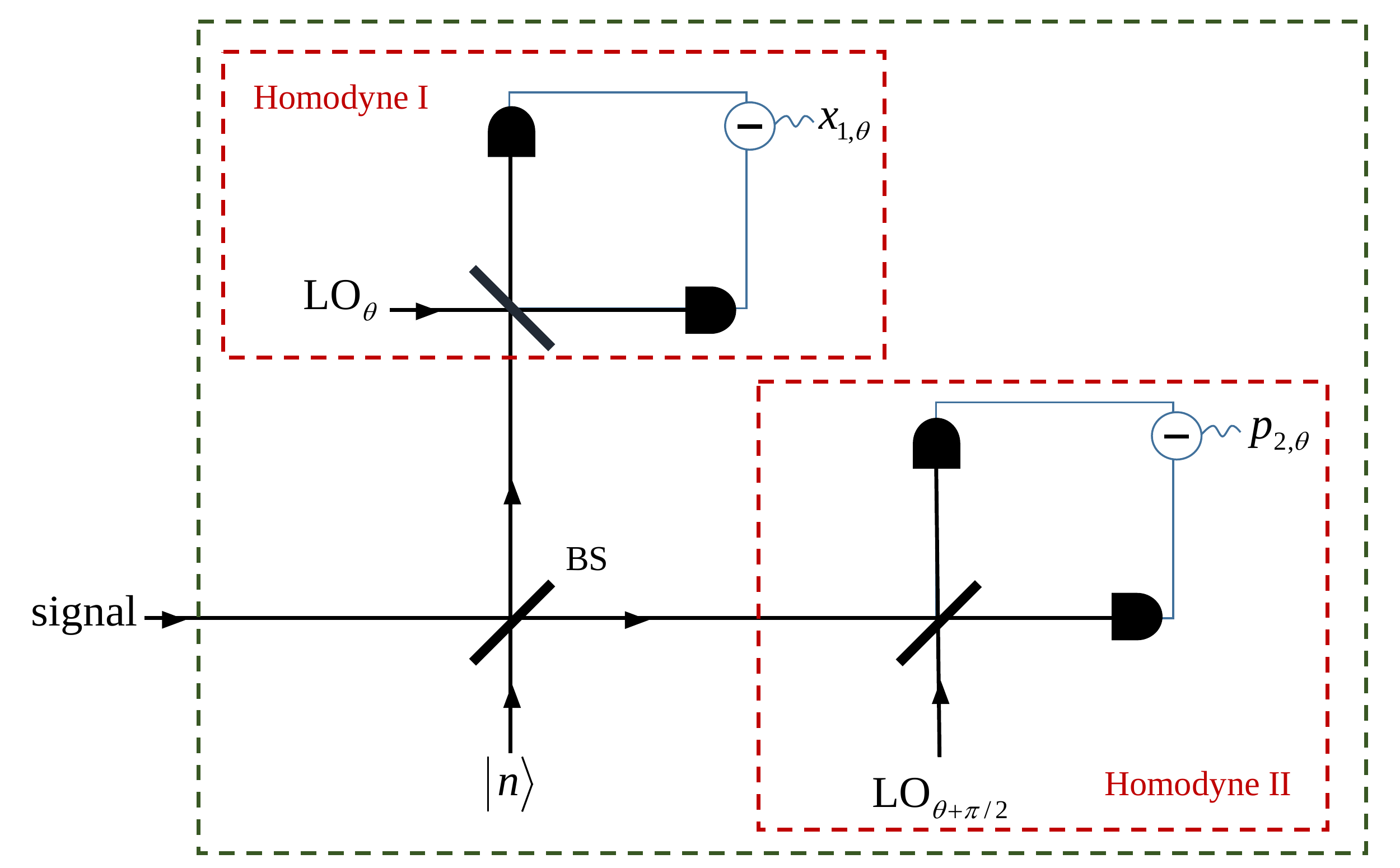}
	\caption{The CV-n measurement device.  The outer green dashed box
	encloses the whole CV-n device, which measures the input signal in the
	displaced Fock state basis.  The inner red boxes represent two
	homodyne detections performed simultaneously that give two CV outcomes
	$x_{1,\theta},p_{2,\theta}$, from the measurement device.} 
	\label{CV-n-meas}
\end{figure}
As depicted in Fig.~\ref{CV-n-meas}, the measurement device works as follows.
The input signal is overlapped on a 50:50 beamsplitter with a number state
$\ket{n}$, and the outputs of the beamsplitter are measured by two conjugate
homodyne detections whose local oscillators have $\pi/2$ phase difference. As
shown in Appendix~\ref{app:cvn}, the POVM elements of this measurement are 
\begin{equation}
	\Pi^n(x_{1,\theta},p_{2,\theta})=\frac{1}{2\pi} D(\alpha) \ketbra{n} D^{\dagger}(\alpha).
\end{equation}
The two real numbers $x_{1,\theta},p_{2,\theta}$, form the results of a simultaneous measurement of two orthogonal quadratures. Notice that for $n=0$, CV-0, we have heterodyne measurement. 

{\it Phase-randomised CV-n (PRCV-n).} If the phase $\theta$ of the local
oscillators is randomised, while the relative phase is fixed, we have PRCV-n
measurement. As discussed in Appendix~\ref{app:cvn}, the outcome of this
measurement is a single non-negative parameter $R=r^2= x_{1,\theta}^2 +
p_{2,\theta}^2$ with the corresponding POVM element
\begin{equation}
\label{povm-phasrand-text}
\Pi^n(R)=n! e^{-R} \sum_{k=0}^{\infty} \frac{R^{k-n}}{k!} \left(L^{k-n}_{n}(R)\right)^2 \ketbra{k}
\end{equation}
with $L^{m}_{k}(.)$ being the generalized Laguerre polynomials.  Though this
measurement has less information, we will concentrate on it as we are only
interested in events where $R \approx 0$ and having the POVM diagonal in the
Fock basis greatly simplifies many calculations.

{\it Discretized-phase-randomised CV-1 (DPRCV-1).} The measurement model
outputs continuous variables (i.e. $R$) and hence these POVMs are describing
probability {\em densities}. To be able to use the methods of~\cite{AA} to
make some definitive statement about computational hardness, one must work in
probabilities and not probability densities.  Hence, we will utilize the
discretized version of the phase-randomized CV-1 (DPRCV-1) measurement through
its ability to distinguish between the single-photon state and other number
states. We divide the range of $0\leq R \leq \infty$ for the POVM elements
in Eq.~(\ref{povm-phasrand-text}) for $n=1$, to two parts $t=\{R|0\leq R \leq
t\}$ which represents the smallest discrete region and $\bar{t}=\{ R|t\leq R
\leq \infty\}$ which represents all the others. Thus, we have two POVM
elements for each interval:
\begin{align}
\label{povm-disc}
\Pi^1_{t}&=\sum_{k=0}^{\infty} \frac{1}{k!} \int_{0}^{t} dR\; e^{-R} R^{k-1 }(k-R)^2 \ketbra{k}\\
&=\sum_{k=0}^{\infty} G(t,k) \ketbra{k},\nonumber
\end{align}
where
\begin{equation}
\label{G-function}
G(t,k)=\frac{1}{k!} \left(k^2 \gamma(k,t)-2k \gamma(1+k,t)+\gamma(2+k,t) \right)
\end{equation}
with $\gamma(k,t)=\int_{0}^{t} dR e^{-R} R^{k-1}$ being the lower incomplete
Gamma function.  As this is a two-outcome measurement and each of the elements is bounded above by $1$ the POVM for all other outcomes can then be written 
\begin{align}
\Pi^1_{\bar{t}}&=I-\Pi^1_{t}.
\end{align}
	
The POVM element in Eq.~(\ref{povm-disc}) can be expanded in a power series as
\begin{align}
&\Pi^1_{t}= \left(\frac{t^2}{2}-\frac{t^3}{3}+O(t^4)\!\right)\!\ketbra{0}\nonumber\\
&+\left(\!t-\frac{3t^2}{2}+\frac{7t^3}{6}+O(t^4)\! \right)\! \ketbra{1}
+ \left(\!t^2-\frac{4t^3}{3}+O(t^4)\!\right)\!\ketbra{2}\nonumber\\
&+\left(\frac{1}{2}t^3+O(t^4)\right)\ketbra{3} 
+O(t^4) \ketbra{4}+\cdots
\end{align}
Therefore, if $t$ is small $t=\epsilon$ such that $\epsilon^2\approx 0$,
the POVM element associated with detecting $R\in[0,\epsilon]$ is
\begin{equation}
\Pi^1_{\epsilon}=\epsilon \ketbra{1}.
\end{equation}
It is this form of the POVM which allows for a measurement that distinguishes
a single photon Fock state from the remainder of the Hilbert space.
	
\begin{figure}[h]
\centering
\includegraphics[width=.9\columnwidth]{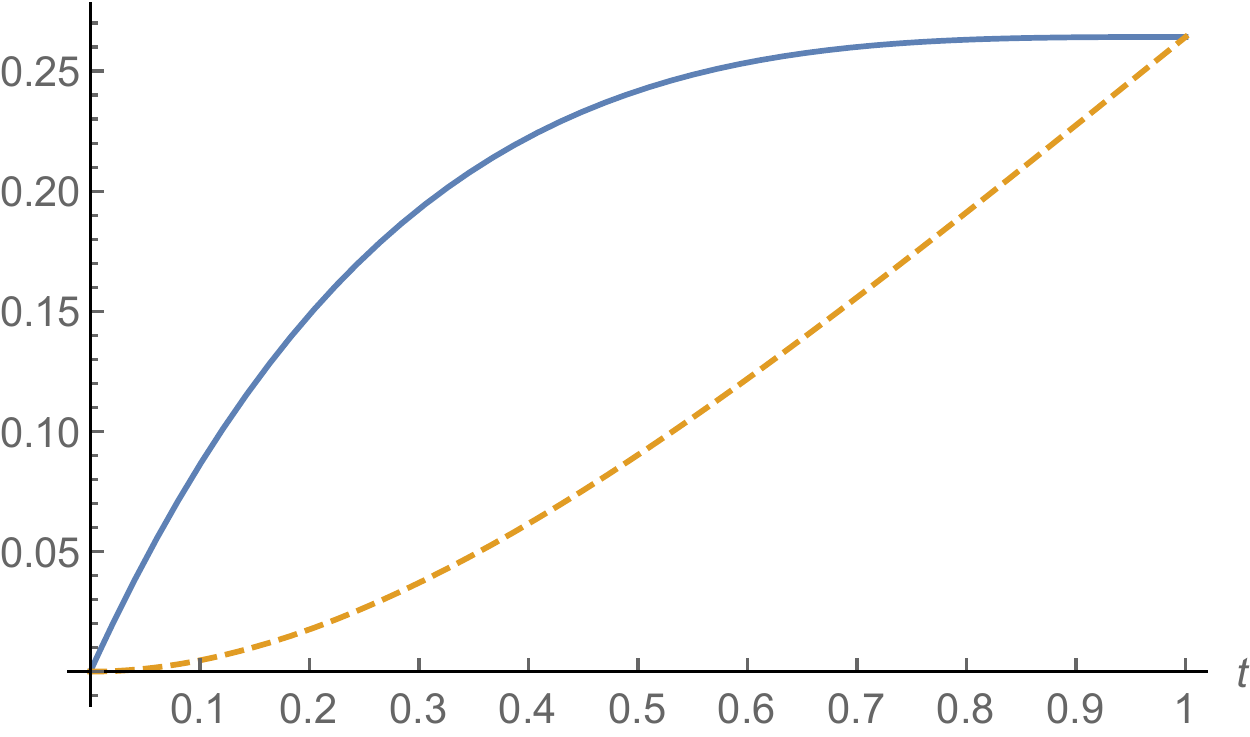}
\caption{Comparison between $\eta(t)$ (solid line) and $p_D(t)$ (dashed line) for the CV-1 detector.} 
\label{etapD}
\end{figure}

To demonstrate what this measurement is detecting consider the case of an
input states being either $\ket{0}$ or $\ket{1}$.  Then when detecting the
single photon state (and for any value of $t$),
\begin{align}
\eta(t)&=\text{Tr}\left[\ketbra{1} \Pi^1_{t} \right]
= 1-e^{-t}(1+t^2)
\end{align} 
can be thought of as the efficiency of the detector. Also
\begin{align}
p_D(t)&=\text{Tr}\left[\ketbra{0} \Pi^1_{t} \right]
= 1-e^{-t}(1+t)
\end{align} 
can be thought as the dark count probability. Fig.~\ref{etapD} compares these
two quantities.

\section{Boson sampling}

We now consider the problem of sampling from the output probability
distribution of a linear-optical network (\LON) using CV-1, PRCV-1, and
DPRCV-1 measurements (where the ancillary Fock state is $n=1$) introduced in
the previous section. We first derive the output probability distribution for
each measurement scheme, and then discuss the complexity of sampling from the
probability distribution.

\subsection{Probability distributions}

\begin{figure}[t]
	\label{BSCV}
	\centering
	\includegraphics[width=.9\columnwidth]{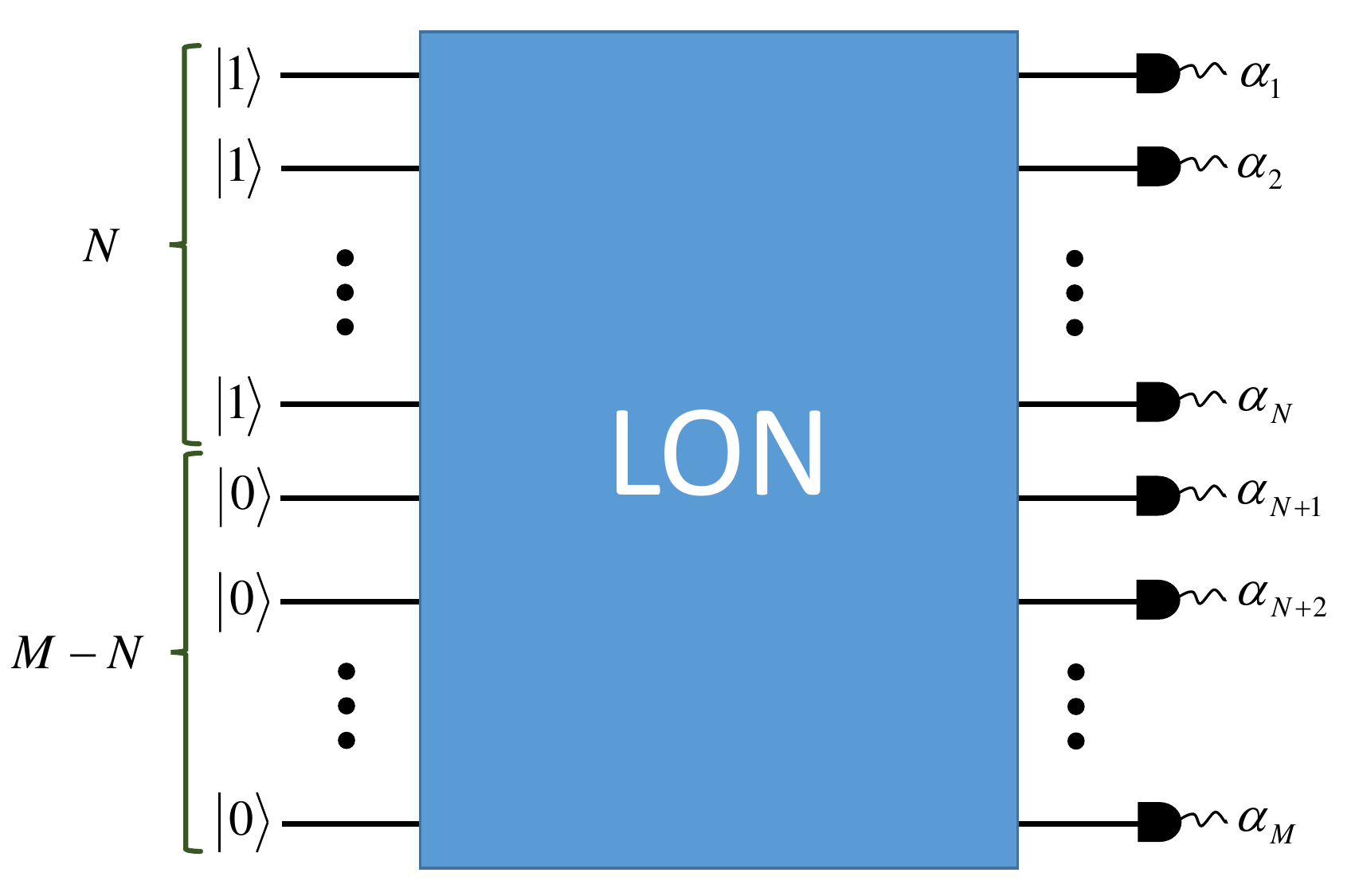}
	\caption{Boson sampling using CV-1 measurements.}
\end{figure}

In this setup, the input state to the \LON\ is
\begin{equation}
\ket{1_N}= \ket{\underbrace{1,1,\cdots,1}_{N},\underbrace{0,0,\cdots,0}_{M-N}}.
\end{equation}
Each output mode is measured by a CV-1 measurement; hence, the overall POVM
elements are
\begin{equation}
\Pi^1(\bm{\alpha})=\frac{1}{(2\pi)^M} D(\bm{\alpha})\ketbra{1_M} D^\dagger(\bm{\alpha}),
\end{equation}
where $\bm{\alpha}=(\alpha_1,\alpha_2,\cdots,\alpha_M)$. As shown in
Fig.~\ref{BSCV}, this setup is equivalent to injecting single photons into a
larger network and performing homodyne measurements at the output, i.e., boson sampling with homodyne measurements.

Therefore, the output probability density is given by
\begin{align}
\label{prob-dis}
P(\bm{\alpha})&= \frac{1}{(2\pi)^M} \left| \bra{1_M} D^\dagger(\bm{\alpha}) \mathcal{U}_{\text{LON}} \ket{1_N} \right|^2\\
&=\frac{1}{(2\pi)^M} \left|\sum_{\bm{n}} \bra{1_M} D^\dagger(\bm{\alpha})\ket{\bm{n}}\bra{\bm{n}} \mathcal{U}_{\text{LON}} \ket{1_N} \right|^2. \nonumber
\end{align}
Here $\mathcal{U}_{\text{LON}}$ is the unitary operation associated with the
\LON, and $\ket{\bm{n}}=\ket{n_1,n_2,\cdots,n_M}$ is the multimode Fock state.
As a \LON\ preserves the number of photons, the sum is restricted to $\nidx$'s
which satisfy $\sum_{i=1}^{M}n_i=N$.

We have \cite{Scheel04}
\begin{equation}
\bra{\bm{n}} \mathcal{U}_{\text{LON}} \ket{1_N} =\text{Per}\left( U_{1_N \times \bm{n}}\right),
\end{equation}
which is the permanent of an $N\times N$ submatrix of the unitary matrix of
\LON, $U_{1_N \times \bm{n}}$ that is corresponding to the first $N$ rows and
columns with multiplicity of $n_i$. The unitary matrix $U$ is defined through
the relation 
\begin{equation}
b^{\dagger}_j=\mathcal{U}_{\text{LON}} a^{\dagger}_{j} \;\mathcal{U}_{\text{LON}}^\dagger=\sum_{i=1}^{M} U_{ij}  a^{\dagger}_{i},
\end{equation}
where $b^{\dagger}_j$ and $a^{\dagger}_{i}$ are modal creation operators for
output $j$ and input $i$, respectively.  Also, using the expression for
$\bra{k}D(\alpha)\ket{m}$ in~\cite{Cah-Glau}, one can simply verify that
\begin{align}
\bra{1_M} D^\dagger(\bm{\alpha})\ket{\bm{n}}
=\prod_{j=1}^{M}e^{-|\alpha_j|^2/2} \frac{(\alpha_{j}^{*})^{n_j-1}}{\sqrt{n_j!}}  (n_j-|\alpha_j|^2).\nonumber
\end{align}  
By using these relations, the output probability density (\ref{prob-dis})
becomes
\begin{equation}
\label{prob-CV}
P(\bm{\alpha})= \frac{e^{-\bm\alpha\bm{\alpha}^\dagger}}{(2\pi)^M} \left|\sum_{\bm{n}} \text{Per} \left(U_{1_N \times \bm{n}}\right) \prod_{j=1}^{M}\frac{(\alpha_{j}^{*})^{n_j-1}}{\sqrt{n_j!}}  (n_j-|\alpha_j|^2) \right|^2\!\!.
\end{equation}
From this expression we can see that if $\alpha_{j}=0$, the probability
density $P(\bm{\alpha})$ is zero unless $n_j=1$. 

We define $\nidx_N=(n_{N,1},n_{N,2},\cdots,n_{N,M})$ as a $M$-tuple whose
elements are either zero or one and the number of ones is $N$, and define
$\bm{\alpha}_{\nidx_N}$ a vector of length $M$ whose element are given by
$[\bm{\alpha}_{\nidx_N}]_j=(1-n_{N,j}) \alpha_j$, i.e., it has $N$ zero
elements corresponding to $n_{N,j}=1$. Using these notations, we can see that
\begin{equation}
P(\bm{\alpha}_{\nidx_N})=\frac{1}{(2\pi)^M} \left| \text{Per} \left(U_{1_N \times \nidx_N}\right)\right|^2 \prod_{n_{N,j}=0} e^{-|\alpha_j|^2} |\alpha_j|^2, 
\end{equation}
where the product is over $M-N$ nonzero elements of $\bm{\alpha}_{\nidx_N}$.
This implies that if $N$ of $M$ $\alpha$'s are zero, the probability
density $P(\bm{\alpha})$ is proportional to the absolute value square of
permanent of a submatix of $U$. For example, for $\nidx_N=1_N$ 
\begin{align}
P(\bm{\alpha}_{1_N})&=P(0,\bm{\alpha}_{M-N})\\
&=\frac{1}{(2\pi)^M} \left| \text{Per} \left(U_{1_N \times 1_N}\right)\right|^2 \prod_{j=N+1}^{M} e^{-|\alpha_j|^2} |\alpha_j|^2.\nonumber
\end{align}

If we use PRCV-1 measurements at the output of \LON, using the POVM
(\ref{povm-phasrand}), the output probability density is given by
\begin{align}
\label{prob-PRCV}
P_{\text{P}}(\bm{R})
= e^{-|\bm{R}|} &\sum_{\bm{n}} \left| \text{Per} \left(U_{1_N \times \bm{n}}\right)\right|^2 
\prod_{j=1}^{M} \frac{1}{n_j!} R_j^{n_j-1} (n_j-R_j)^2,
\end{align}
where $\bm{R}=(R_1,R_2,\cdots,R_M)$ and $|\bm{R}|=R_1+R_2+\cdots+R_M$.
Defining $\bm{R}_{\nidx_N}$ similar to $\bm{\alpha}_{\nidx_N}$, we have
\begin{align}
P_{\text{P}}(\bm{R}_{\nidx_N})=& \left| \text{Per} \left(U_{1_N \times \bm{n}_N}\right)\right|^2\\
&\times \prod_{n_{N j}=0} \frac{1}{n_j!} e^{-R_j} R_j^{n_j-1} (n_j-R_j)^2.\nonumber
\end{align}

For DPRCV-1 measurements at the output, the probability of detecting outcome $\midx_{N}$, that is, $N$ clicks within interval $t$ and $M-N$ clicks within $\bar{t}$, using Eq.~(\ref{G-function}), is given by
\begin{align}
\label{prob-DPRCV}
P_{\text{D}}(\midx_{N})= &\sum_{\bm{n}} \left| \text{Per}(U_{1_N \times \bm{n}})\right|^2\\
&\times \prod_{m_{N j}=1} G(t,n_j) \prod_{m_{N j}=0} (1-G(t,n_j)). \nonumber
\end{align}
If $t$ is very small, this probability can be expanded to leading order in a
powers of $t$ as
\begin{align}
\label{prob-expand}
P_{\text{D}}(\midx_{N})=& 
\left| \text{Per} \left(U_{1_N \times \midx_{N}}\right)\right|^2 
\left(t^N + O(t^{N+1}) \right)\\
&+\sum_{\midx=\midx_{N}^1} |\text{Per}\left(U_{1_N \times \bm{m}}\right)|^2 
O(t^{N+2}),\nonumber
\end{align}
where $\midx_{N}^1$ are $M$-tuples whose elements are equal to $\midx_{N}$
except one 0 and one 1 are interchanged.  

Notice that the probability distribution (\ref{prob-DPRCV}) corresponding
to DPRCV-1 measurements is coarse-grained version of probability density
(\ref{prob-PRCV}), and that itself is phase randomised version of
(\ref{prob-CV}). Therefore, if exact sampling from the probability distribution
(\ref{prob-DPRCV}) is classically hard, exact sampling from the other
probability distributions must be classically hard as well. In the following
two subsections we discuss exact and approximate sampling from the probability
distribution (\ref{prob-DPRCV}).

\subsection{Exact boson sampling}

To leading order in $t$, Eq.~(\ref{prob-expand}) give a distribution for which
hardness of exact sampling can be determined as we shall show in this section.
Merely setting the $O(t^{N+1})$ terms to zero results in a probability
\begin{equation}
P_{\text{D}}(1_{N}) =  \Usubprob t^N.
\end{equation}
A probability of this form does not pose any problems for the argument of
hardness in exact sampling.  However, it is important to consider higher order
terms as exact sampling allows for sampling using any matrix even those with
vanishingly small permanents.  For example, one could consider the case where
the permanent $\Usubprob = O(t^2)$ but the unwanted higher order terms have
$\Uprob{\Usm} = O(1)$.  Furthermore, as $\Uprob{\Usm}$ defines a probability
distribution over all $\bm{n}$ (i.e. the $\BosonSampling$ distribution) we
have
\begin{equation}
	\sum_{\midx=\midx_{N}^1} |\text{Per}\left(U_{1_N \times \bm{m}}\right)|^2  \leq 1.
\end{equation}
This means that, using Eq.~(\ref{prob-expand}), the probability can be written as
\begin{equation}
	\label{probability}
	P_{\text{D}}(1_{N}) =  t^N \left(\left(1-O(t)\right) \Usubprob + E\right).
\end{equation}
where $E = O(t^2)$ is an ``error'' over the desired probability.
Unfortunately $E$ contributes to the probability in an additive sense.
However, as we have used the upper bound on the matrix permanents that
contribute to $E$ we know that this bound is independent of $N$.  So if we
choose $E$ to be a small constant, we need to quadratically vary $t$ to
achieve that constant.  Furthermore, with constant $E$, the sizes of
sub-matrix permanents that can be estimated must be lower bounded with a bound
that depends on $E$.

To make this more explicit consider the case where $0 < L \leq \Usubprob \leq
1$.  If we used the approximate counting algorithm using $P_{\text{D}}(1_{N})$
from Eq.~(\ref{probability}) then we would have an estimate $\tilde{p}$ which
satisfies
\begin{widetext}
\begin{equation}
	(\Usubprob + E) g^{-1} < \tilde{p} < (\Usubprob + E) g
\end{equation}
where $g>1$ is the multiplicative error factor from the estimation.  We can
also write
\begin{equation}
	(\Usubprob - |E|) g^{-1} < \tilde{p} < (\Usubprob + |E|) g
\end{equation}
and using $\Usubprob > L$
\begin{equation}
	\Usubprob( 1 - |E|/L) g^{-1} < \tilde{p} < \Usubprob( 1 + |E|/L) g
\end{equation}
Now if $|E|/L < 1/2$ (arbitrarily, could be any constant less than 1), we have
\begin{equation}
	\Usubprob ( 1 + 2 |E|/L)^{-1} g^{-1} < \tilde{p} < \Usubprob( 1 + 2 |E|/L) g
\end{equation}
where we now have a multiplicative error estimate with error 
$g^\prime=(1+2|E|/L)g$.
\end{widetext}

The \#P-hardness of approximating $\text{Per}(X)^2$ (Theorem 28 from \cite{AA}) where
$X$ is an $N \times N$ matrix, requires a choice of $g$ that is polynomially
dependent on $N$.  This will be achieved here if the extra $(1+2|E|/L)$ term
maintains a polynomial scaling.  

At this point, we consider the origin of the added error term $E$ to argue how
$L$ must scale in terms of $N$.  All of the schemes described in Section III
depend on a continuous output parameter based on the CV-1 style measurement.
However, any experimental realisation will output values to within some
precision.  One particular way to do this, with a connection to discrete
computational outputs is to consider a discretisation of the results into
$b$-bit integers.  The outcomes $R$ from PRCV-1 are non-negative, so in this
case consider $b$-bit non-negative integers.  $R$ is unbounded for large
values, but in our scheme we are only interested in small values so we can
choose an arbitrary fixed boundary above which all values are assigned the
same bitstring output.  With these requirements, results can be partitioned
into $2^b$ regions such that the largest result corresponds to all outcomes
above some fixed value, say $1$.  In the simplest case, if $b=1$ then we would
have the one bit binary string with `$0$' representing all results of the
continuous value $R$ between $0$ and $1$ and the binary string `$1$'
representing values greater than $1$.  Increasing $b$ and continuing this
division of the results, we find that the range of values covered by equal
size partitions of the values between $0$ and $1$ we have $t = 2/(2^b-1) =
O(2^{-b})$.  It is the scaling in $b$, and not $t$, that is required to be
considered when analysing the complexity of the problems modelled on this
device.

We will proceed with the analysis below assuming this simple partitioning of
the values of $R$.  It should be noted that other discretisation strategies
can be considered.  For example, a change of the discretisation thresholds
such that smaller regions around zero can be considered by using a common
technique called ``companding''.  However, one must be careful here not to
sacrifice too much probability by having the zero region scale
super-exponentially.  Any scaling which results in the region at zero scaling
as $O(2^{-poly(b)})$ is sufficient for the exact sampling argument below.
This is due to the nature of the Stockmeyer approximate counting algorithm
utilized in AA.  As these strategies do not change the complexity hardness
result we present, we will not consider them further in this paper.

Now we will return to considering the error term $|E|$.  This quantity, as
stated above, scales as $O(t^2)$.  But in the number of bits used for
discretisation it will scale as as $O(2^{-2b})$. To achieve a $poly(N)$
scaling of $|E|/L$ a scaling in $L$ of $O(2^{-2b})$ would counteract the error
leaving a $O(1)$ overhead. 

So our procedure will achieve a multiplicative estimate with a very quickly
decaying lower bound $L$ on the matrix permanents that can estimate.  But
what actually determines the choice of the lower bound $L$ that is permitted?
The proof of the \#P-hardness for approximating $\text{Per}(X)^2$ proceeds
in~\cite{AA} by using this estimation polynomially many times to compute
$\text{Per}(X)$ given that $X$ was a zero-one matrix.  As part of the proof
one needs to use the estimation procedure in a binary search towards a matrix
whose permanent is zero and the procedure stops when sufficient precision has
been achieved to determine the permanent of the zero-one matrix, as it must be
an integer value.  Introducing a lower bound on the permanents for which the
estimation is valid would have to be compatible with this final precision.

Never-the-less, in the hardness proof for approximate $\BosonSampling$, it is
assumed that the input matrices $X$ have matrix elements that have a Gaussian
distribution.  If one is to accept the conjectures of~\cite{AA}, then there is
a low probability of randomly having $\text{Per}(X)$ near zero.  Specifically, using
the words from~\cite{AA}: 
\begin{quote}
	\ldots if $X \sim \mathcal{G}^{n \times n}$ is Gaussian then \ldots a
	$1-1/poly(n)$ fraction of [$\text{Per}(X)$'s] probability mass is
	greater than $\sqrt{n!}/poly(n)$ in absolute value, $\sqrt{n!}$ being
	the standard deviation.
\end{quote} 
Using this we could choose the lower bound of $Per(X)$ to scale as 
$\sqrt{N!}/poly(N)$.  However, we have written the bound for our construction
in terms of the submatrix of the unitary matrix $U$.  Following the embedding
procedure from~\cite{AA} this reduces the size of $X$ by a factor involving
the matrix norm $||X||$.   Therefore the probability reduces by a factor of
$||X||^{2N} \leq 2^{poly(N)}$ as $N$ (the number of input photons) is a proxy
for the problem input size (i.e. the matrix $X$) which must be efficiently
represented in $N$.  This results in a scaling of $L = O( ||X||^{-2N}
N!/poly(N))$.  In terms of the discretisation bits the scaling is $b = O(N
\log N + poly(N))$ where the $poly(N)$ in this scaling is from the polynomial
which bounds $||X||^{-2N}$.  So a polynomial scaling in the size of the
discretisation allows for approximation of permanents which are highly
probable when using an approximate $\BosonSampling$ algorithm.  This argument
shows that the exact sampling problem presented here, which permits
multiplicative permanent estimation with permanents with an exponentially
decreasing lower bound, must also be hard to compute with classical resources. 

\subsection{Approximate sampling}

The arguments just made do not permit one to continue the hardness
argument through to the case of approximate sampling as was done in~\cite{AA}
for the CV distributions we have constructed above.  The reason is quite
simple, only an exponentially small subset of events (i.e. detections near the
origin) are used to make the exact sampling argument.  Conversely, for
the same reason, one also cannot definitively conclude that approximate
sampling from this distribution is an efficient classical task.

The approximate sampling criterion for this CV distribution would be 
\begin{equation}
	\int |P(\bm{\alpha}) - Q(\bm{\alpha})| d \vec{\alpha} < \beta
\end{equation}
where $P$ is the probability density for our distribution of CV events from
the device described above and $Q$ is the computable approximation.  The event
used for sampling above is the set of $\bm{\alpha}$, where $N$ $\alpha$'s are
around a ball of radius $t$ around the origin.  Hence all of this error could
potentially be concentrated on our event.  This would dominate any exponential
pre-factors of the matrix permanent in the probability exponentially reducing
the signal that is being fed into the approximate counting algorithm.

Approximate sampling is shown in~\cite{AA} by potentially utilising all
possible (or more precisely the collision-free subspace of) events to perform
the approximate counting.  The algorithm cannot know which event is being
probed, so a concentrate of error as described above would result in the
approximate sampling algorithm being an exact sampling algorithm for almost
all results.

So what is required in the CV case is more events.  To show classical
hardness for this distribution, one needs to argue that events away from the
origin are also \#P-hard when used exactly.  This is difficult for this
construction as these events do not reduce to Fock basis measurements under
any approximation.  Also, there must be combinatorially enough events so that
the approximate sampling error can be considered low enough on average over
all the events so that hardness is maintained.  Even if this specific
criterion cannot be met, this does not mean that the distribution is efficient
for a classical computation.  To show classical efficiency, either a
constructive proof demonstrating efficiency (e.g. the methodology used
in~\cite{SRK16-suff}) or some other proof technique ruling out classical
hardness would be required.

\section{Conclusion}

We have constructed a continuous variable measurement model we call CV-n which
measures in a displaced number state basis.  The measurement model can be
achieved by mixing a $n$ photon state with the input on a 50:50 beam splitter
then making homodyne measurements on the output.  We have shown how a $n$ Fock
basis state measurement can be approximately achieved using this
measurement utilising those cases when the homodyne measurement outcomes are
simultaneously small.

We then showed that this measurement model is compatible with the exact
$\BosonSampling$ problem, a computing task that has been shown to be inefficient
for any classical device to simulate.  We have discussed how this model as
presented here is not compatible with approximate $\BosonSampling$ as the
detection events utilised, when compared with the whole event space, is too
small.

\section*{Acknowledgements}

This work was supported by the Australian Research Council Centre of
Excellence for Quantum Computation and Communications Technology (Project No.
CE110001027).

\appendix

\begin{widetext}

\section{CV-n measurement: Continuous variable measurement in the displaced number state}
\label{app:cvn}

In this appendix we obtain the POVM elements of the CV-n measurement; see Fig. 1.  The
POVM elements of this measurement are of this form 
\begin{equation}
\label{povm-psi}
\Pi^n(x_{1,\theta},p_{2,\theta})=\frac{1}{c}\ketbra{\Psi(x_{1,\theta},p_{2,\theta})},
\end{equation}
where $x_{1,\theta}$ and $p_{2,\theta}$ are the outcomes of the fist and the second homodyne measurements, respectively, and $c$ is the normalization constant such that $\int dx_{1,\theta}\; dp_{2,\theta} \Pi(x_{1,\theta},p_{2,\theta})=I$. 

The POVM elements of the first homodyne measurement are 
\begin{equation}
\label{povm-1hom}
\ketbra{x_{1,\theta}}=e^{-i \hat{P}_{1,\theta} x_{1,\theta}}\ketbra{x_{1,\theta}=0}e^{i \hat{P}_{1,\theta}x_{1,\theta}},
\end{equation}
where 
\begin{equation}
\hat{P}_{1,\theta}= -\hat{X}_1 \sin{\theta}+\hat{P}_1 \cos{\theta}.
\end{equation}
Notice that $\ket{x_{1,\theta}=0}$ is an infinitely squeezed vacuum state, and can be written as
\begin{equation}
\label{infsq-1}
\ket{x_{1,\theta}=0}=\lim_{r\rightarrow\infty} \hat{S}(r e^{i2\theta})\ket{0}
\end{equation} 
with $\hat{S}(\xi)=\exp\left[(\xi^* \hat{a}^2 -\xi \hat{a}^{\dagger 2})/2 \right]$ being the squeezing operator and $\ket{0}$ being the vacuum state. 

For the second homodyne measurement we have
\begin{equation}
\label{povm-2hom}
\ketbra{p_{2,\theta}}=e^{i \hat{X}_{2,\theta} p_{2,\theta}}\ketbra{p_{2,\theta}=0}e^{-i \hat{X}_{2,\theta}p_{2,\theta}},
\end{equation}
where 
\begin{equation}
\hat{X}_{2,\theta}= \hat{X}_2 \cos{\theta}+\hat{P}_2 \sin{\theta}.
\end{equation}
Similarly, we have
\begin{equation}
\label{intsq-2}
\ket{p_{2,\theta}=0}= \lim_{r\rightarrow\infty} \hat{S}(-r e^{i2\theta})\ket{0}.
\end{equation}
\\

Now using Eqs.~(\ref{povm-1hom}) and (\ref{povm-2hom}), $\ket{\Psi(x_{1,\theta},p_{2,\theta})}$ in Eq.~(\ref{povm-psi}) is given by
\begin{equation}
\label{Psi}
\ket{\Psi(x_{1,\theta},p_{2,\theta})}= \frac{1}{\mathcal{N}}\bra{n} \underbrace{\hat{\mathcal{U}}_{\text{BS}} e^{-i \hat{P}_{1,\theta} x_{1,\theta}} e^{i \hat{X}_{2,\theta} p_{2,\theta}}}_{A} \ket{x_{1,\theta}=0,p_{2,\theta}=0},
\end{equation}
where $\mathcal{N}$ is the normalization constant. Using the following relations for the 50:50 beamsplitter
\begin{align}
\hat{\mathcal{U}}_{\text{BS}}
\begin{pmatrix}
\hat{X}_1 \\
\hat{X}_2
\end{pmatrix}
\hat{\mathcal{U}}_{\text{BS}}^\dagger= 
\begin{pmatrix}
\frac{1}{\sqrt{2}} & \frac{-1}{\sqrt{2}}\\
\frac{1}{\sqrt{2}} & \frac{1}{\sqrt{2}}
\end{pmatrix}
\begin{pmatrix}
\hat{X}_1 \\
\hat{X}_2
\end{pmatrix} \\
\hat{\mathcal{U}}_{\text{BS}}
\begin{pmatrix}
\hat{P}_1 \\
\hat{P}_2
\end{pmatrix}
\hat{\mathcal{U}}_{\text{BS}}^\dagger= 
\begin{pmatrix}
\frac{1}{\sqrt{2}} & \frac{-1}{\sqrt{2}}\\
\frac{1}{\sqrt{2}} & \frac{1}{\sqrt{2}}
\end{pmatrix}
\begin{pmatrix}
\hat{P}_1 \\
\hat{P}_2
\end{pmatrix},
\end{align}
the expression $A$ in Eq.~(\ref{Psi}) becomes
\begin{equation}
\hat{\mathcal{U}}_{\text{BS}} e^{-i \hat{P}_{1,\theta} x_{1,\theta}} e^{i \hat{X}_{2,\theta} p_{2,\theta}}=D_1(\beta_1) D_2(\beta_2) \hat{\mathcal{U}}_{\text{BS}},
\end{equation}
where
\begin{align}
\beta_1&=\frac{1}{2} \left(x_{1,\theta}\cos{\theta}-p_{2,\theta}\sin{\theta}\right)+i \frac{1}{2}\left(x_{1,\theta} \sin{\theta}+p_{2,\theta}\cos \theta \right) \\
\beta_2&=\frac{1}{2} \left( -x_{1,\theta}\cos{\theta}-p_{2,\theta}\sin{\theta}\right)+i\frac{1}{2} \left(-x_{1,\theta} \sin{\theta}+p_{2,\theta}\cos \theta\right),
\end{align}
and $D(\beta)$ is the displacement operator. Thus, Eq.~(\ref{Psi}) becomes
\begin{equation}
\label{Psi2}
\ket{\Psi(x_{1,\theta},p_{2,\theta})}= \frac{1}{\mathcal{N}}\bra{n} D_1(\beta_1) D_2(\beta_2) \hat{\mathcal{U}}_{\text{BS}}  \ket{x_{1,\theta}=0,p_{2,\theta}=0}.
\end{equation}
Using
\begin{align}
\hat{\mathcal{U}}_{\text{BS}} \ket{x_{1,\theta}=0,p_{2,\theta}=0}&=\lim_{r\rightarrow\infty} \hat{\mathcal{U}}_{\text{BS}} \hat{S}(re^{i2\theta})\hat{S}(-re^{i2\theta})\ket{0,0}\\
&=\lim_{r\rightarrow\infty} \frac{1}{\cosh r}\sum_{k=0}^{\infty} \left( e^{2i\theta} \tanh r\right)^k \ket{k,k}\\
&=\lim_{r\rightarrow\infty} \frac{e^{2i\theta \hat{n}_2} }{\cosh r}\sum_{k=0}^{\infty} \left(\tanh r\right)^k \ket{k,k},
\end{align}
we can write Eq.~(\ref{Psi2}) as
\begin{equation}
\label{Psi3}
\ket{\Psi(x_{1,\theta},p_{2,\theta})}= \frac{1}{\mathcal{N}} D_1(\beta_1) \lim_{r\rightarrow\infty} \frac{1}{\cosh r}\sum_{k=0}^{\infty} \left(\tanh r\right)^k \ket{k} \bra{n} D_2(\beta_2) e^{2i\theta \hat{n}_2} \ket{k}.
\end{equation}
It can be shown that
\begin{equation}
\label{inter-n-k}
\bra{n} D_2(\beta_2) e^{2i\theta \hat{n}_2} \ket{k}= e^{2i\theta n} \bra{n} D_2(\beta_2')  \ket{k}=e^{2i\theta n} \bra{k} D_2(-\beta_2'^*)  \ket{n},
\end{equation}
where 
\begin{equation}
\beta_2'=
\frac{1}{2} \left(-x_{1,\theta}\cos{\theta}+p_{2,\theta}\sin{\theta}\right)+i \frac{1}{2}\left(x_{1,\theta} \sin{\theta}+p_{2,\theta}\cos \theta \right),
\end{equation}
and the second equality can be seen using this expression
\begin{equation}
\label{disp-matr}
\bra{n} D(\alpha)  \ket{k}=\sqrt{\frac{k!}{n!}} e^{-|\alpha|^2/2} \alpha^{n-k} L^{n-k}_{k}(|\alpha|^2)
\end{equation}
with $L^{m}_{k}(.)$ being the generalized Laguerre polynomials. 

By substituting from Eq.~(\ref{inter-n-k}) into Eq.~(\ref{Psi3}), we get
\begin{equation}
\label{Psi4}
\ket{\Psi(x_{1,\theta},p_{2,\theta})}= \frac{1}{\mathcal{N}} D_1(\beta_1) e^{2i\theta n} \lim_{r\rightarrow\infty} \frac{1}{\cosh r}\sum_{k=0}^{\infty} \left(\tanh r\right)^k \ket{k} \bra{k} D_2(-\beta_2'^*)  \ket{n}.
\end{equation} 
As this state must be normalized in the limit of $r\rightarrow \infty$, it is straightforward to see that
\begin{equation}
\mathcal{N}^{-1}=\cosh r.
\end{equation}

As $\lim_{r\rightarrow\infty} \tanh r=1$, we now have
\begin{align}
\label{Psi5}
\ket{\Psi(x_{1,\theta},p_{2,\theta})}&= e^{2i\theta n} D_1(\beta_1)  \left( \lim_{r\rightarrow\infty} \sum_{k=0}^{\infty} \left(\tanh r\right)^k \ketbra{k}\right) D_2(-\beta_2'^*)  \ket{n}\\
&=e^{2i\theta n} D_1(\beta_1) D_2(-\beta_2'^*) \ket{n} \\
&= e^{2i\theta n} e^{(-\beta_1\beta_2'+\beta_1^*\beta_2'^*)/2} D(\beta_1-\beta_2'^*) \ket{n}\\
&= e^{2i\theta n} e^{(-\beta_1\beta_2'+\beta_1^*\beta_2'^*)/2} D(\alpha) \ket{n},
\end{align}
where
\begin{equation}
\alpha= (x_{1,\theta}+i p_{2,\theta}) e^{i\theta}= r e^{i\theta+i\phi}
\end{equation}
\\

Therefore, POVM elements of the CV-n measurement are
\begin{equation}
\label{povm-psi2}
	\Pi^n(x_{1,\theta},p_{2,\theta})=\frac{1}{2\pi} D(\alpha) \ketbra{n} D^{\dagger}(\alpha).
\end{equation}

Notice that, it can be simply checked that
\begin{equation}
\frac{1}{2\pi} \int d^2\;\alpha D(\alpha) \ketbra{n} D^{\dagger}(\alpha) =I.
\end{equation}

\subsection{Phase randomized CV-n measurement}
If the phase $\theta$ of the local oscillator is randomized, with the relative phase $\pi/2$ being fixed, we have
\begin{equation}
\Pi^n(R)=\frac{1}{2\pi}\int d\theta D(\sqrt{R} e^{i\theta+i\phi})\ketbra{n}D^\dagger (\sqrt{R} e^{i\theta+i\phi}),
\end{equation}
where 
\begin{equation}
R=r^2= x_{1,\theta}^2 + p_{2,\theta}^2.
\end{equation}

The matrix elements of this operator in the Fock basis are
\begin{align}
\bra{k} \Pi^n(R) \ket{l}&=\frac{1}{2\pi}\int d\theta \bra{k} D(\sqrt{R} e^{i\theta+i\phi})\ketbra{n}D^\dagger (\sqrt{R} e^{i\theta+i\phi}) \ket{l}\\
&=\frac{1}{2\pi} e^{-R} \frac{n!}{\sqrt{k! l!}} (\sqrt{R})^{k+l-2n} L^{l-n}_{n}(R) L^{k-n}_{n}(R) \int d\theta e^{i(\theta+\phi)(k-l)}\\
&=\frac{1}{2\pi} e^{-R} \frac{n!}{\sqrt{k! l!}} (\sqrt{R})^{k+l-2n} L^{l-n}_{n}(R) L^{k-n}_{n}(R) 2 \pi \delta_{k,l}.
\end{align}

Therefore, the POVM elements of the phase randomized measurement are given by
\begin{equation}
\label{povm-phasrand}
\Pi^n(R)=n! e^{-R} \sum_{k=0}^{\infty} \frac{R^{k-n}}{k!} \left(L^{k-n}_{n}(R)\right)^2 \ketbra{k}.
\end{equation} 

It can be simply verified that
\begin{equation}
\int_{0}^{\infty} dR\; \Pi^n(R) =I.
\end{equation}
\\

For $n=1$, using
\begin{equation}
L^{k-n}_{n}(R)=\sum_{j=0}^{n} \frac{(-1)^j k!}{j! (k-n+j)! (n-j)!} R^j,
\end{equation}
we have
\begin{equation}
L^{k-1}_{1}(R)=k-R;
\end{equation}
so the POVM elements (\ref{povm-phasrand}) becomes
\begin{equation}
\Pi^1(R)= e^{-R} \sum_{k=0}^{\infty} \frac{R^{k-1}}{k!} \left(k-R\right)^2 \ketbra{k}.
\end{equation}
Notice that
\begin{equation}
\Pi^1(0)=\ketbra{1},
\end{equation}
as expected.
\end{widetext}

\end{document}